\documentclass[conference]{IEEEtran}
\IEEEoverridecommandlockouts
\usepackage{cite}
\usepackage{amsmath,amssymb,amsfonts}
\usepackage{algorithmic}
\usepackage{graphicx}
\usepackage{textcomp}
\usepackage{diagbox}
\usepackage{xcolor}
\usepackage{tikz}
\usepackage{algorithm}
\usepackage{float} 
\usepackage{subcaption}
\usepackage{silence}

\usetikzlibrary{shapes.geometric, arrows}
\def\BibTeX{{\rm B\kern-.05em{\sc i\kern-.025em b}\kern-.08em
    T\kern-.1667em\lower.7ex\hbox{E}\kern-.125emX}}

\begin{document}
\title{NLP-based Cross-Layer 5G Vulnerabilities Detection via Fuzzing Generated Run-Time Profiling\\
}

\author{\IEEEauthorblockN{Zhuzhu Wang}
\IEEEauthorblockA{\textit{School of Systems and Enterprises} \\
\textit{Stevens institute of technology}\\
Hoboken, USA \\
zwang326@stevens.edu}
\and
\IEEEauthorblockN{Ying Wang}
\IEEEauthorblockA{\textit{School of Systems and Enterprises} \\
\textit{Stevens institute of technology}\\
Hoboken, USA \\
ywang6@stevens.edu}

}
\maketitle

\begin{abstract}
The effectiveness and efficiency of 5G software stack vulnerability and unintended behavior detection are essential for 5G assurance, especially for its applications in critical infrastructures. Scalability and automation are the main challenges in testing approaches and cybersecurity research. In this paper, we propose an innovative approach for automatically detecting vulnerabilities, unintended emergent behaviors, and performance degradation in 5G stacks via run-time profiling documents corresponding to fuzz testing in code repositories. Piloting on srsRAN, we map the run-time profiling via Logging Information (LogInfo) generated by fuzzing test to a high dimensional metric space first and then construct feature spaces based on their timestamp information. Lastly, we further leverage machine learning-based classification algorithms, including Logistic Regression, K-Nearest Neighbors, and Random Forest to categorize the impacts on performance and security attributes. The performance of the proposed approach has high accuracy, ranging from $ 93.4 \% $ to $ 95.9 \% $, in detecting the fuzzing impacts. In addition, the proof of concept could identify and prioritize real-time vulnerabilities on 5G infrastructures and critical applications in various verticals.

\end{abstract}


\section{Introduction}
The fifth generation (5G) cellular network holds promise for achieving the vision of universal connectivity. It enables various application scenarios and verticals,  including connected vehicles, remote robotic surgery, etc., which has a profound impact on society\cite{Hussain2019}. Meanwhile, there is a paradigm shift in the core network architecture of telecommunication systems in the 5G networks, which have transitioned to software-defined infrastructure and thereby reduced reliance on hardware-based network functions. This change has dramatically improved network efficiency, performance, and robustness, however, it has also made the network more vulnerable, as software systems are often more susceptible to disruption than hardware systems. Therefore, rigorous testing against vulnerabilities becomes critical in 5G and future G systems \cite{9701880}. Security concerns, connection failures, and performance latency have often been a topic of interest to academia and industry. 

Fuzzing technology is an effective statistic-based testing technique in detecting unknown vulnerabilities of a product or protocol by automating the generation and execution of a large number of random test cases. Many researchers used fuzzy testing techniques to check the security of communication protocols, including Packet Forwarding Control Protocol[PFCP]\cite{9790271}, 5G Non-Access-Stratum[NAS] protocol\cite{9868872}, etc. However, existing research on fuzz testing suffers from challenges in utilizing fuzz testing in network protocols and software implementations: scalability of search space.  

The exponential increase of fuzzing testing case corresponding to the search space leads to the challenge of providing a complete picture of the entire security threat or vulnerability with mere fuzzing and raise insurmountable difficulties to the computation complexity when search space scales. To address the scalability issue in fuzz testing search space, formal verification uses mathematical models to precisely describe a system. Recently, Megha et al.\cite{9652163} used an automatic verification tool called ProVerif to verify the 5G Extensible Authentication Protocol and Authentication and Key Agreement[5G-EAP-AKA] protocol. Jingjing et al.\cite{8923409} based on the Scyther model checker to study the 5G Extensible Authentication Protocol and Transport Layer Security[5G-EAP-TLS] protocol. However, these methods rely on an in-depth understanding of the 5G specifications and require high time and labor costs to build the mathematical model.

To reduce challenges in computation complexity and reliance on human expertise in specific projects for vulnerability detection, our objective in this study is to semantically predict vulnerabilities based on the profiling traces of various 5G platforms.  We propose an innovative approach for detecting vulnerabilities and unintended emergent behaviors in general 5G stacks automatically based on NLP models with fuzz testing profiling via LogInfo. The method uses semantic information for vulnerability detection, which not only predicts potential detection before the occurrence of failures and vulnerability but also breaks the application scenario barrels and allows for an implementation-independent and cross-platform solution. The proposed solution can be applied to different protocols as well as software stacks.

Event logging records detailed run-time information for a system, which supports engineers in understanding system behavior and tracking potential problems. The log file is the text data generated by the printout code embedded in the program by the program developer to assist in debugging, which is used to record the variable information and program execution status when the program is running, etc. And therefore log messages are a valuable resource used for many tasks, such as performance monitoring \cite{zhang2012research} and anomaly detection \cite{sipola2011anomaly}. It can locate the specific log and event information and also locate the exception request instance. Based on the NLP technology, calculating the center distance according to the time interval, we use log files to successfully detect if there is a vulnerability under the fuzzing test.

The contributions of this paper are summarized as follows:
\begin{itemize}

\item An NLP-based LogInfo vulnerability detection solution for 5G fuzz testing is designed and implemented. The BERT-based model is trained for sentence-to-vector, and the weighed periodic distances are calculated to determine vulnerabilities and unintended emergent behaviors.  

\item The design solution discovers insights for potential vulnerabilities and unintended emergent behaviors via the pattern comparison between fuzzed testing and standardized testing. With the featured scalability and automation, it enables zero-day attack resistance and enhances real-time sustainability.

\item Machine-learning-based classification models are developed for classifying the type of vulnerabilities. We also demonstrate its practical value of light-weighted and high accuracy. 
\end{itemize}

The remainder of this paper is organized as follows. Section \uppercase\expandafter{\romannumeral2} summarizes the related work. Section    \uppercase\expandafter{\romannumeral3} describes the necessary preliminary knowledge of the 5G protocol stack. 
In section \uppercase\expandafter{\romannumeral4}, we present the design and implementation of the proposed approach. Section \uppercase\expandafter{\romannumeral5} discusses and analyzes the experimental results.
Section \uppercase\expandafter{\romannumeral6} concludes the paper and discusses the future work plan, respectively.

\section{Related work}
Previous work in cellular network security includes both formal verification \cite{edris2020formal} \cite{zhang2019formal} and fuzzy testing \cite{hu2021fuzzing}\cite{he2022intelligent}. A formal model is constructed and security properties are extracted to formally validate the model. Formal verification can be effective in detecting vulnerabilities in protocol designs. Manual methods of analysis and verification Manual methods of vulnerability analysis and verification have produced some attacks, but it requires extensive experience from security researchers and lacks universality. Fuzzing has also been applied to security test analysis of protocols, but it requires manual sample construction and is a semi-automatic approach. In \cite{9519388}, the author utilizes NLP as a guide to discover vulnerabilities in LTE specifications. In addition, many studies have used text analysis techniques to automatically discover various errors in different domains, such as SSL/TLS implementations\cite{10.1007/11736790_9}, API Misuse Detection \cite{10.1145/3372297.3423360}. 

Different from the above works, our research provides a systematic framework to discover vulnerabilities from real log files, which is an expression of the specific implementation of the protocol. As detailed guidance on the implementation of protocols is still lacking, code flaws can introduce new security risks at the implementation level of 5G protocols. Especially for lack of testing for unexpected data during the implementation of 5G protocol stacks, which can lead to potential security risks not being detected promptly, thus affecting the robustness of the protocol implementation. By analyzing the related log files, we can identify vulnerabilities and propose solutions promptly.

\section{BACKGROUND}
\subsection{5G protocol stack}
A protocol stack is a prescribed hierarchy of software layers, and the 5G protocol stack between the UE and eNB shown in Fig. 1 can be divided into three layers. Starting from the bottom layer, these layers are divided into the physical layer, the data link layer, and the network layer.
\begin{itemize}
\item Physical layer: The physical layer transmits information through the air interface between the UE and the gNB.
\item The data link layer is responsible for connecting the physical layer to the network layer and is divided into three parts into three subparts. Starting from the bottom layer, these subparts are the Media Access Control (MAC) protocol, the Radio Link Control (RLC) protocol, and the Packet Data Convergence Protocol (PDCP).
\item Network layer: The network layer, similar to the data link layer, is divided into three subparts. The subcomponents of the network layer are the non-access layer (NAS), the radio resource control (RRC), and the Internet protocol (IP). Among them, NAS is divided into Mobility Management (NAS-MM) and NAS Session Management (NASSM), which interact with Access and Mobility Function (AMF) and Session Management Function (SMF), respectively.
\end{itemize}

The log files used for this experiment are based on the work in \cite{JingdaYang20235GListen-and-Learn}, and piloting RRC protocols in the srsRAN\cite{SoftwareRadioSystems2021SrsRANSRS} platform. RRC protocols are designed as one of the critical cellular protocols for radio resource management, which not only provides the upper layers with wireless resource parameters from the network system but also control the main parameters and behavior of the lower layers.

\begin{figure}[htbp]
    \centering
    \centerline{\includegraphics[width=9cm]{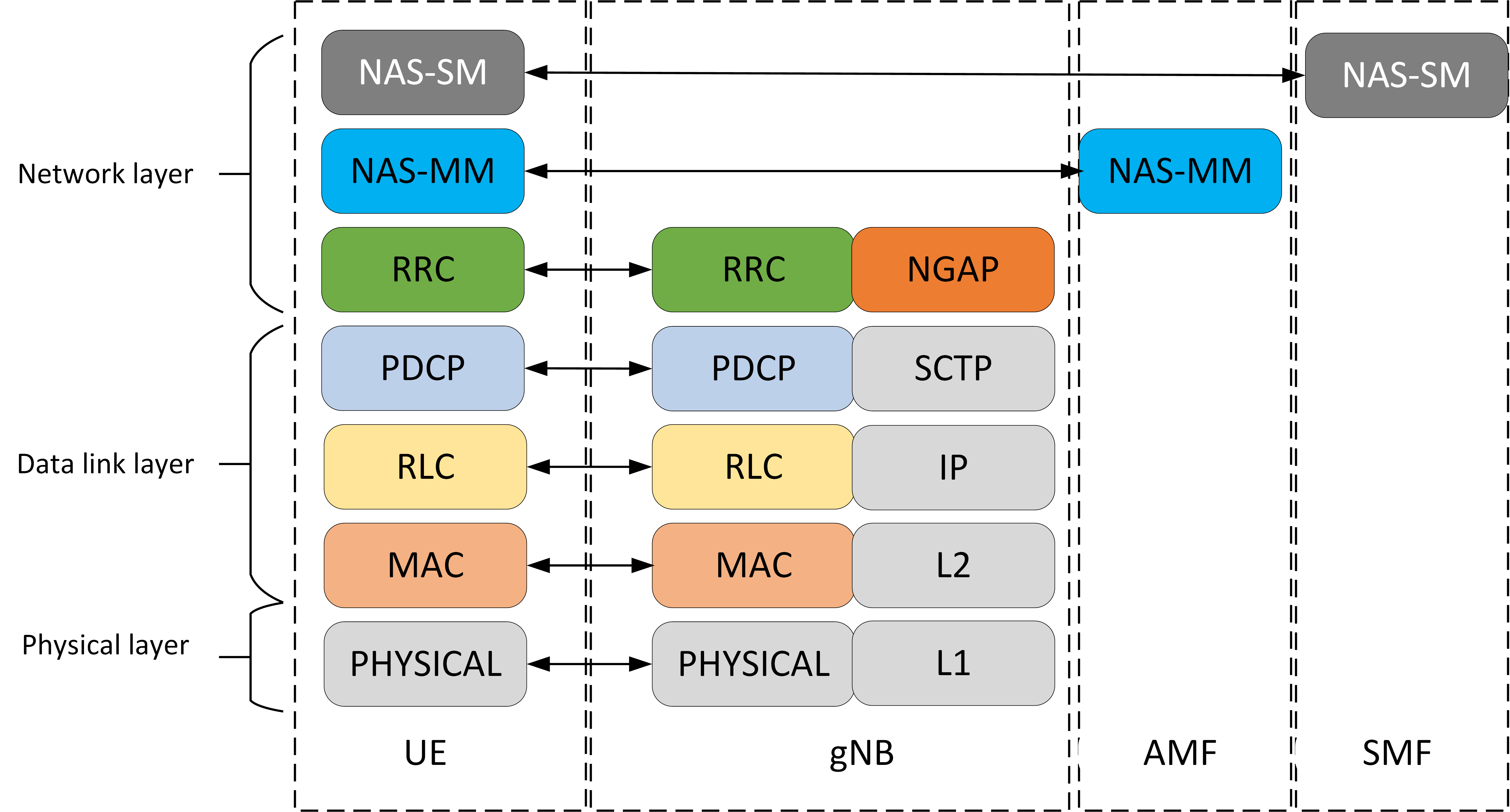}}
    \caption{5G Stack Architecture.}
\end{figure}

\section{The proposed approach}
\subsection{System description}
Fig. 1 illustrates a typical description of 5G architecture. There are three main entities (roles) and we explain them below.

\begin{figure}[htbp]
    \centering
    \centerline{\includegraphics[width=9cm]{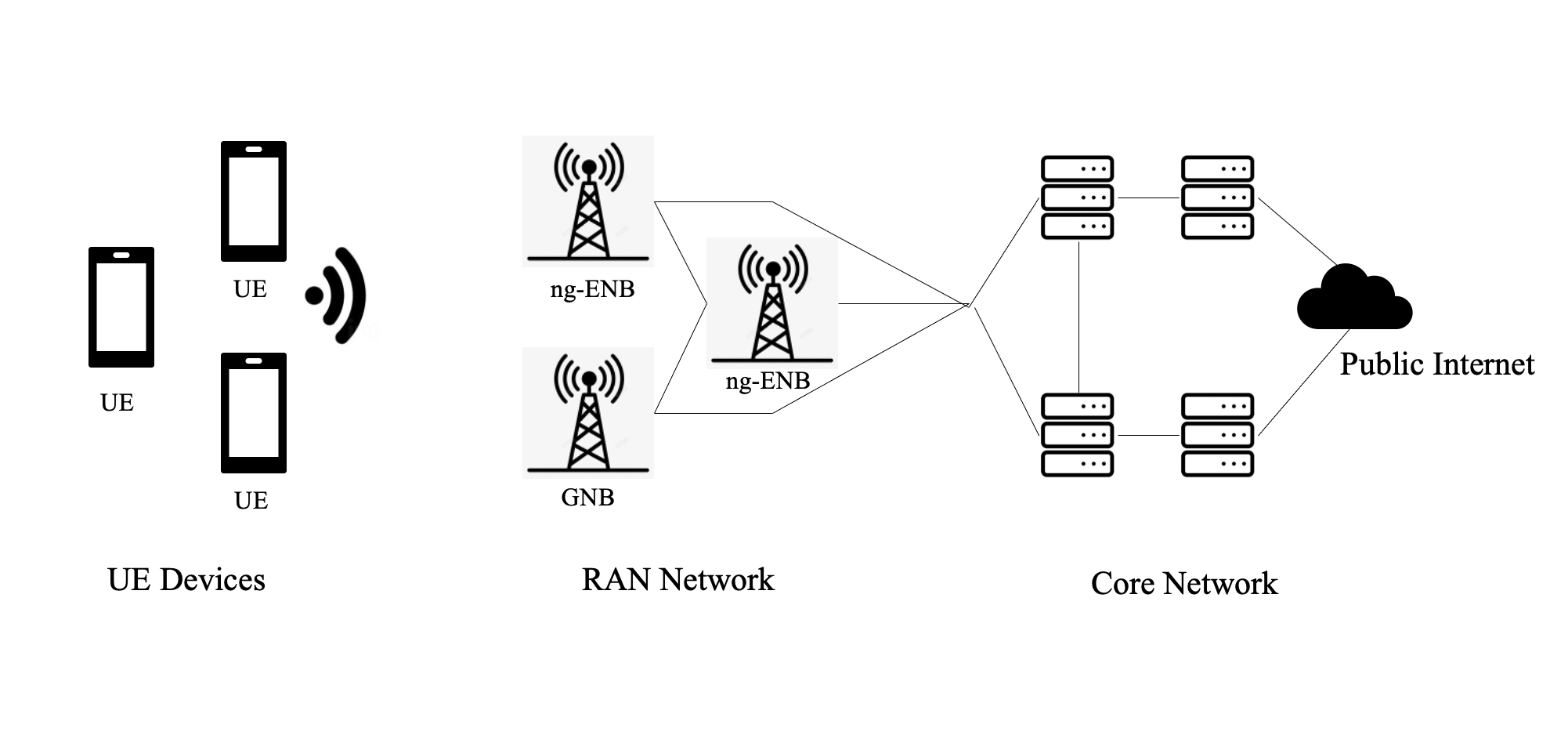}}
    \caption{5G architecture.}
\end{figure}

\begin{itemize}
\item User Equipment (UE): represents a subscriber’s mobile device, such as a cell phone, tablet, or modem.
\item Radio access network (RAN): represents the telecommunications network that connects UE to the core network via a radio connection. There are two main forms: Next Generation e-NodeB (ng-eNB) and Next Generation NodeB (gNB). The gNB allows 5G UE to connect with the 5G NG(New Generation) core using the 5G NR(New Radio)air interface.
\item Core network (CN): represents the network that provides services to mobile subscribers through the RAN. It also serves as a gateway to other networks, such as to the public switched telephone network or the public cloud.
\end{itemize}

\subsection{Approach overview}
Fig. 3 shows the designed 5G stacks' unintended behaviors and vulnerability detection via fuzzing-generated run-time profiling based on NLP. Here, we used command-level fuzzing to detect vulnerabilities. Specifically, exchanged legitimate messages will be observed and collected to the fuzzing message candidate pool, and at the command level, commands are replaced by other commands in the same physical channel to test whether any communication error state occurs. The detailed running profile will be recorded in the log files. 

The proposed method is shown in Fig. 3 and mainly consists of 4 parts: generating and collecting log files, natural language processing, dimensional reduction, and classification. Below is a detailed explanation of each step:
\begin{figure*}[htbp]
    \centering
\centerline{\includegraphics[width=18cm]{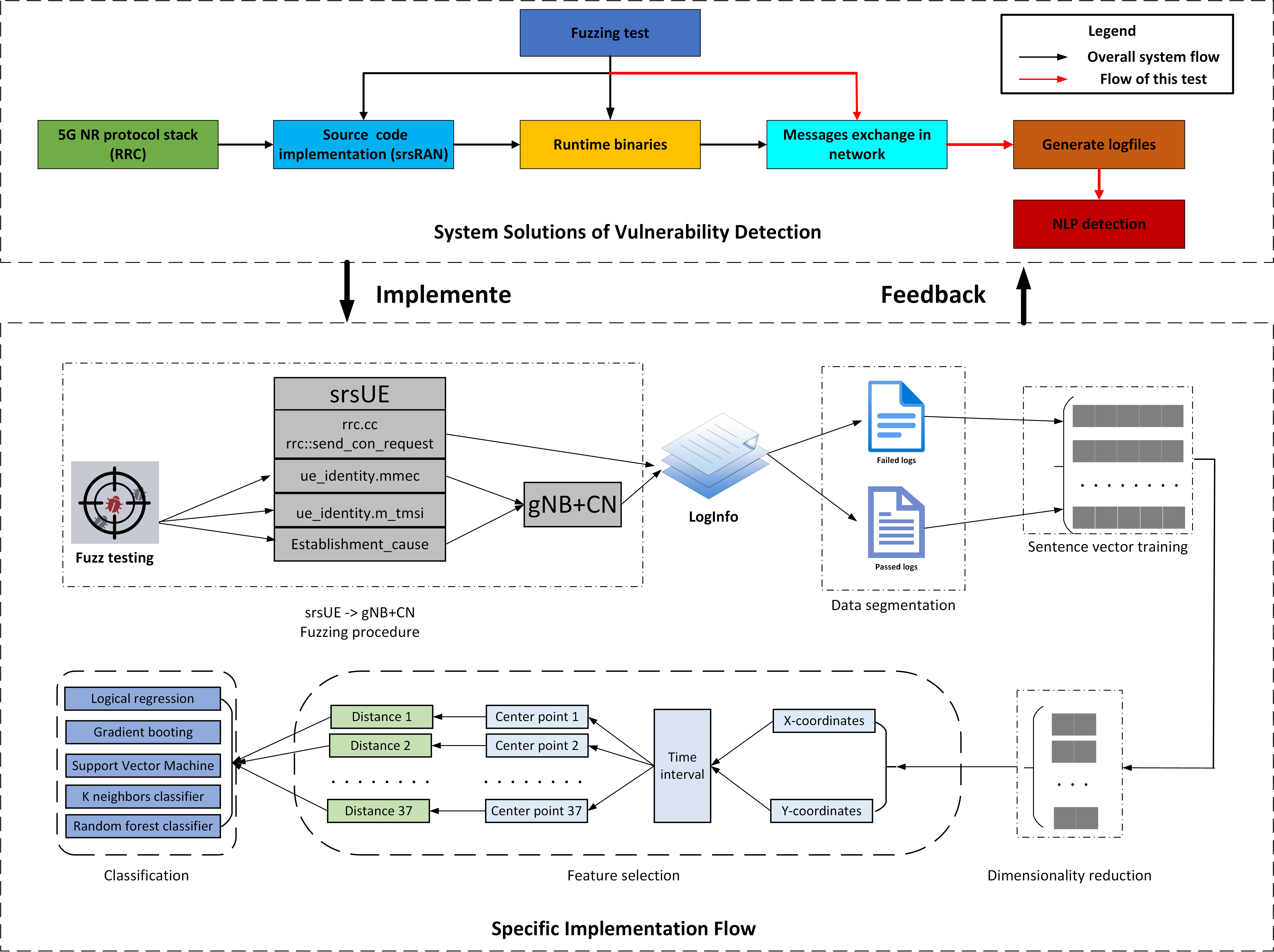}}
    \caption{System design and specific experimental protocols.}
\end{figure*}

\subsubsection{Data collection}We fuzz the RRC protocols and tunneled Non-Access Stratum(NAS) protocols through message recorder, injection, and modification.   
The srsRAN system is monitored using packet capture software and logs are generated using packet capture software. When srsRAN executes, the gNB is initiated, and its respective outputs are sent to a log file. According to whether the communication was successfully achieved, we divide the logs into two parts: successfully connected and failed connected. 
\subsubsection{Natural language processing}Log files are semi-structured language, we could not use them directly. And therefore, we choose the sentence-BERT model to obtain the sentence vector. Specifically, the log files are pre-processed and input into the model for training to obtain a semantic representation of the text: the sentence vector. 
\subsubsection{Dimensional reduction}To avoid dimensional disasters and speed up the training process, we use the t-SNE model to perform down-scaling operations.
\subsubsection{Classification} Without losing generality, we chose common classification algorithms as our model, such as Logistic Regression, KNeighbors, Random Forest, etc.

\subsection{Experimental setup}
\subsubsection{\textbf{Profiling Acquiring via Fuzz Testing}}Fuzzy testing is the generation and injection of unexpected inputs called "fuzzy" inputs, by slightly changing a fundamental change and introducing each change into the system to observe its effect. Here, the srsUE executable has been modified to replace the field values in the message of interest with the fuzzy values of the current test case. Then, the new fuzzy message is sent from the UE to the gNodeB(gNB). During the fuzzing, the recorded states include the following information: ‘message time’, ‘original bytes’, ‘RRC channel’, ‘message type’, and ‘physical channel’. When the monitor in gNB detects the \textit{rrcConnectionSetupComplete} message, the testing case will be terminated and labeled as a successful connection. When the monitor in gNB cannot detect the \textit{rrcConnectionSetupComplete} message within a predefined timeout limit (600 seconds in the proof of concept experiments), it is considered a failed connection. The inclusion of this keyword indicates that the RRC connection is completed and communication can proceed. Profiling files that do not include this specific keyword indicate an effective fuzzing that leads to an unsuccessful connection. By annotating the profiling files with a search for the keyword \textit{rrcConnectionSetupComplete}, all files are labeled as either success ($=1$) or fail ($=0$).

However, relying on keyword searching results is a fragile approach that can be misleading in case of changes to the keyword expression or format changes in the profiling. Thus, we use keyword searching as the ground truth and propose a semantic-based robust detection mechanism. 

The binary annotation provides critical information on the effectiveness or impact of fuzzing cases. However, annotating a case as successful ($=1$) is not necessarily an indicator that the fuzzing did not have an impact on communication. The latency generated by the fuzzing could lead to severe Denial of Service in scaled cases with a large number of users and significantly reduce the system's capacity. To address this, we also annotate the profiling with the duration of the RRC connection.

Log messages in the profiling files take the format like this: \textit {Timestamp \quad[Layer ] \quad Level\quad Content} (e.g., \textit{17:52:25.246 \quad [RLC ] \quad Info\quad DRB1 Tx SDU}), among them, PHY-layer log messages have some additional details: \textit {Timestamp \quad[Layer] \quad Level\quad [Subframe]\quad Channel:\quad Content} (e.g., \textit{17:52:26.094\quad [PHY1]\quad Info\quad [05788]\quad PDSCH:\quad l\_crb=1, harq=0, snr= 22.1 dB, CW0: tbs=55, mcs=22, rv=0, crc=OK, it=1, dec\_time=12 us}). 

The srsENB profiling files could be configured by log levels to display each layer of the network stack correspondingly. Contents of each profiling file may include the time stamp of each step, and log information for semantic information abstraction. It details the steps in the srsRAN platform enabling hex message output. We adapt our models to run on various log levels, including debug, info, warning, error, and none. A visual example of the srsENB profiling file is shown in Fig.4.

\begin{figure}[htbp]
    \centering
    \centerline{\includegraphics[width=9cm]{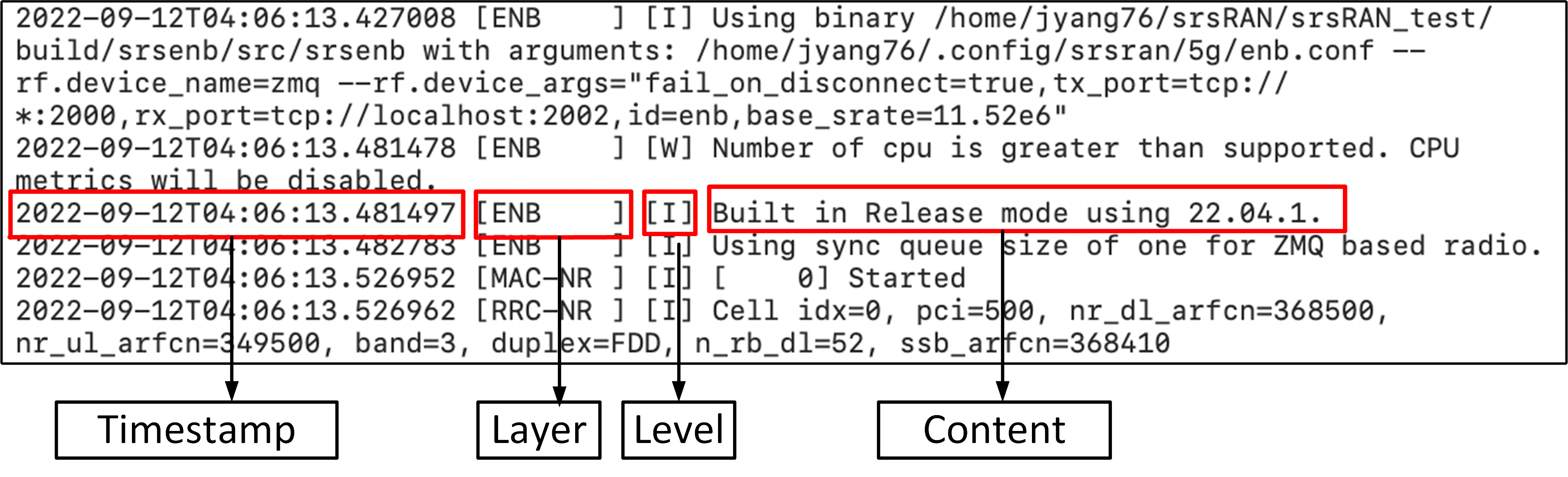}}
    \caption{Example of srsENB events.}
\end{figure}

\subsubsection{\textbf{Data pre-processing}}As shown in Fig.4, each of the log files contains a varying number of lines and each line includes a different amount of words, including non-alphanumeric characters. Firstly, we use regular expressions to remove these non-alphanumeric characters. It should be emphasized that the sentences input to the subsequent model here are those LogInfo belonging to the same timestamp. In addition, for the timestamp information, we extract it and save it separately but not as a corpus for training.

\subsubsection{\textbf{Definition and calculation of distance}}As raw logs are often semi-structured textual messages, they need to be converted into digital features firstly. And therefore, we choose natural language processing tools to map every log event to matrices. The aim is to learn semantic information about the logs \cite{chen2021experience}. Here, we adopt the Bidirectional Encoder Representations from the Transformers (BERT) model. BERT is a useful tool for calculating text similarity, which is published by researchers at Google AI Language \cite{devlin2018bert}. The BERT model leverages the attention mechanism in the transformer, training methods of the mask language model, and word order information encoded in the embedding to implement the simultaneous use of contextual and word order information. The BERT model can be trained by combining the contexts in all layers of the neural network to obtain contextually relevant bi-directional feature representations. The trained model is thus able to understand the semantics in conjunction with the context.

In the actual training process, we mainly refer to\cite{reimers2019sentence} and use\cite{reimers-2020-multilingual-sentence-bert} as a pre-training model for our corpus training. 

\begin{algorithm}[h]
\caption{Sentenc2vec based on sentence-BERT \& t-SNE}
\begin{algorithmic}[1]
\REQUIRE{ $N_{th}$ log file's sentences $\chi_i $ = $\{s_1^N,s_2^N,...,s_n^N\}$ after pre-processing}
\ENSURE{Coordinate mapping for each sentence: $\textbf{Y}^N$}
\STATE For $N_{th}$ log file, given a set of input sentences $\chi_i $ = $\{s_1^N,s_2^N,...,s_n^N\}$
\STATE Training these sentences to obtain sentence vectors: 

$[\textbf{x}_1^N,\textbf{x}_2^N,...,\textbf{x}_n^N]$ (via sentence-BERT)
\FOR{each $i,j \in [1,n]$}
\IF{$i\ne j$}
\STATE $ p_{j|i} = \frac{exp(-||\textbf{x}_i^n-\textbf{x}_j^n||^2/2\delta_i^2)}{\sum_{k\neq i}exp(-||\textbf{x}_k^n-\textbf{x}_i^n||^2/2\delta_i^2)}$;
\ELSE
\STATE set $ p_{i|i} = 0 $
\ENDIF
\STATE $p_{ij} = \frac{p_{j|i} + p_{i|j}}{2N}$;
\ENDFOR
\STATE Initialize $2$-dimensional vectors [$ \textbf{y}_1^N$, ..., $\textbf{y}_n^N $] based on $\mathcal{N}(0, 10^{-4}\textit{I})$
\FOR{each $i,j \in [1,n]$}
\IF{$i\ne j$}
\STATE $ q_{j|i} = \frac{(1+||\textbf{y}_i-\textbf{y}_j||^2)^{-1}}{\sum_{k\neq l}(1+||\textbf{y}_k-\textbf{y}_l||^2)^{-1}} $;
\ELSE
\STATE set $ q_{i|i} = 0 $;
\ENDIF
\STATE $q_{ij} = \frac{q_{j|i} + q_{i|j}}{2N}$;
\ENDFOR
\STATE Given learning rate $ \eta $ and momentum $ \alpha(t) $
\STATE Gradient optimization formula: 

$\frac{\delta C}{\delta y_i} = 4\sum_j(p_{ij}-q_{ij})(y_i-y_j)(1+||{y}_i-{y}_j||^2)^{-1} $;
\REPEAT
\STATE $\textbf{y}_{i+1}^N = \textbf{y}_i^N + \eta \frac{dC}{dY} + \alpha(t)(\textbf{y}_{i-1}^N - \textbf{y}_{i-2}^N) $;
\UNTIL Minimize: $ KL(P||Q) = \sum\limits_{i}\sum\limits_{j}p_{ij}\log\frac{p_{ij}}{q_{ij}} $
\end{algorithmic}
\end{algorithm}

After the training of the BERT model, we obtain a sentence vector of dimension 512. To facilitate the computation and reduce memory, the extraction of valid information, and the rejection of useless information, we reduce the dimensionality of the sentence vector. The T-Distribution Stochastic Neighbour Embedding(t-SNE) was chosen for the dimensionality reduction algorithm. The t-SNE is a machine learning algorithm for dimension reduction, which visualizes high-dimensional data by giving the position of each data point on two maps, proposed by Laurens van der Maaten and Geoffrey Hinton in 2008 \cite{van2008visualizing}. As a nonlinear dimensionality reduction algorithm, t-SNE is very suitable for visualizing high-dimensional data down to 2 or 3 dimensions, by using a small distance to generate a large gradient for dissimilar points to repel them. But it is not infinitely large to avoid dissimilar issues being too far away.
Algorithm 1 shows the specific vector training and dimensionality reduction process. 

After inputting the log files into algorithm 1, we get a 2-dimensional coordinates vector for each sentence from LogInfo. For each log event, there is a unique timestamp corresponding to it. And therefore, we calculate the time duration from starting time to each log event that occurred. The accuracy of the calculation is measured in seconds. Under this condition, each second interval will contain multiple log events. For every running time interval, we calculate the corresponding center point. Here, it is important to note that because the first 4 seconds are the time that trying to build a connection, and therefore, the values of $T$ are ${0,4,5,6,...,40}$. Then, we calculate the distance of every point. The specific calculation formula is shown in Algorithm 2.

\begin{algorithm}[h]
\caption{Features' value calculation method based on center distance}
\begin{algorithmic}[1]
\REQUIRE{$N_{th}$ log file's 2-dimensional vectors $\textbf{Y}^N $ = $\{y_t^1,y_t^2\}$ based on the algorithm 1}
\ENSURE{Every second's ($t$) distance of central point 
$ {Distance}_{t} $}
\STATE Initialize time interval $T = \{0,4,...,40\}$ 
\FOR{each $t \in T$}
\STATE $\bar{y}_{t}^1 = \frac{\sum_{i=1}^{n}y_{i}^{1}}{n}$;\
\STATE $\bar{y}_{t}^2 = \frac{\sum_{i=1}^{n}y_{i}^{2}}{n}$
\ENDFOR
\FOR{each $t \in T$}
\STATE Calculate distance: ${Distance}_{t} = \sqrt{\bar{y}_{t}^{1}+\bar{y}^{2}_{t}}$;
\ENDFOR
\end{algorithmic}
\end{algorithm}

To validate our approach, we select mainstream binary classification algorithms as our training model. These classification algorithms are provided by \textit{scikit-learn} and are very easy to select and train, including Support Vector Machine, k-Nearest Neighbors, Logistic Regression, etc.

\begin{figure}[htbp]
    \centering
    \centerline{\includegraphics[width=9cm]{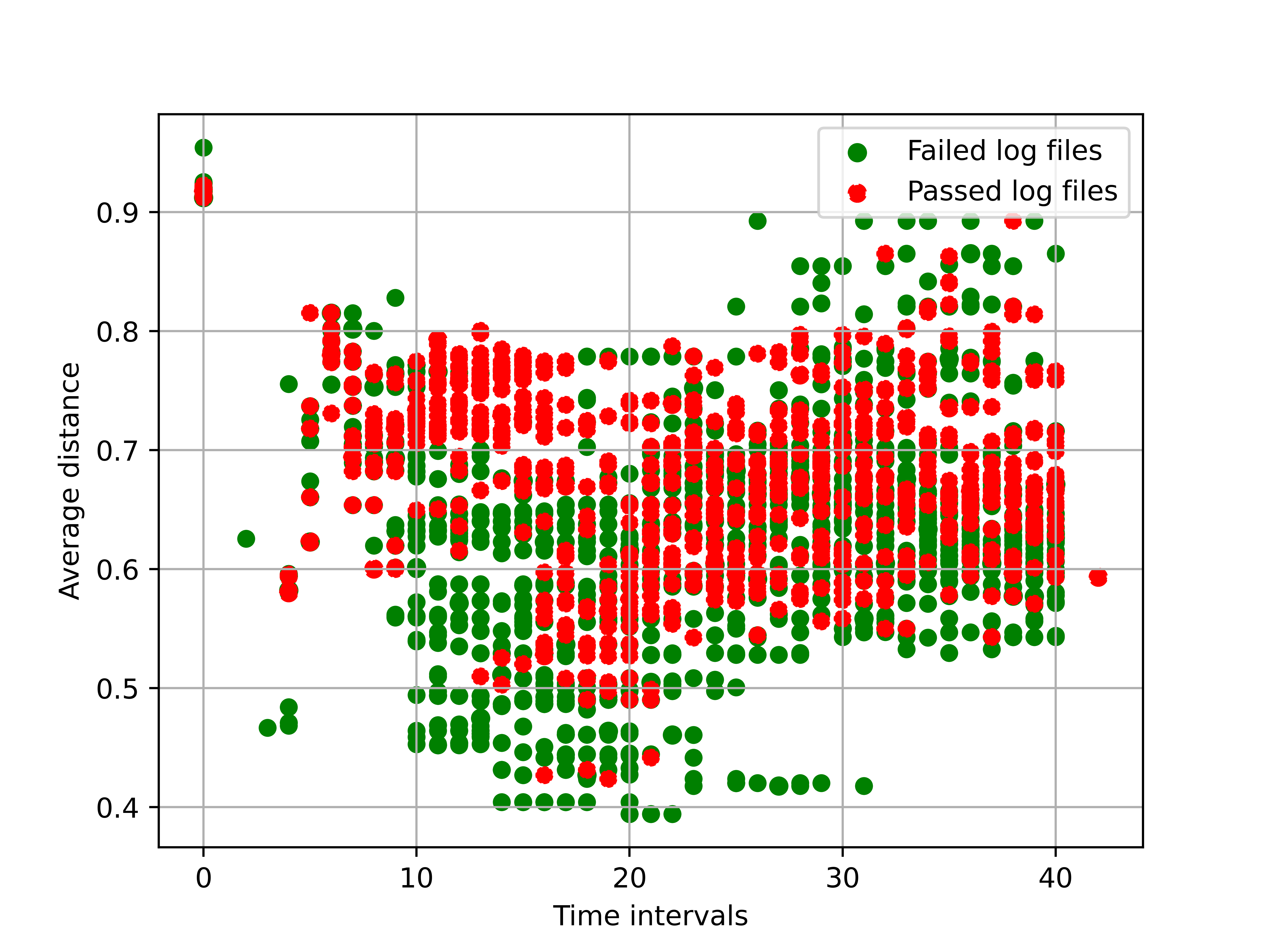}}
    \caption{The visualization of the partial log files}
\end{figure}

\section{Experimental results and analysis}
In this section, we test the validity of our model with real log files. As mentioned earlier, these log files were generated by fuzzing tests applied to the 5G radio resource control (RRC) protocol piloting on srsRAN. Fig. 5 presents the result of the partial log files after distance calculation. Every point represents the distance from the center of all event logs generated during this second to the origin when the system runs to this second.

Fig. 6 presents the Receiver Operating Characteristics(ROC) and Area Under the ROC(AUC) of different classifiers. Here, we choose five common binary classification algorithms from sci-kit-learn as our classification models. 
\begin{figure}[htbp]
    \centering
    \centerline{\includegraphics[width=9cm]{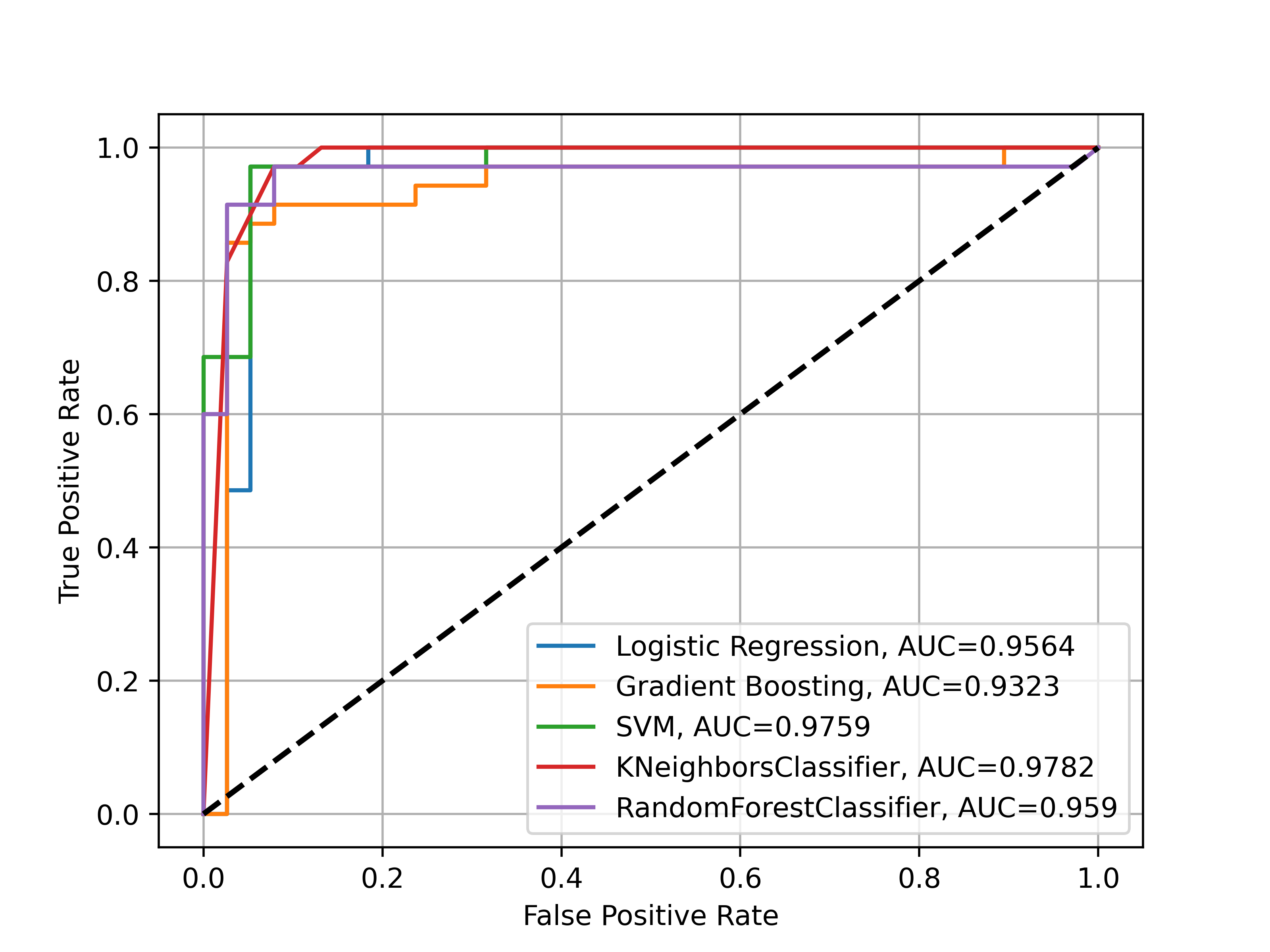}}
    \caption{ROC and AUC results corresponding to different algorithms.}
\end{figure}

Through further analysis, we found that the main difference between failed log files and passed log files occurs between the 10th and 17th seconds as shown in Fig. 7, which means we do not need to use the whole log file to input the prediction model. As shown in Fig. 8, we generate the accuracy distributions using different time intervals and find that only takes until the tenth second for the accuracy rate to reach a high level, which means we could determine in advance if the connection is successful.

\begin{figure}[htbp]
	\begin{minipage}{0.49\textwidth}
		\vspace{3pt}	\centerline{\includegraphics[width=\textwidth]{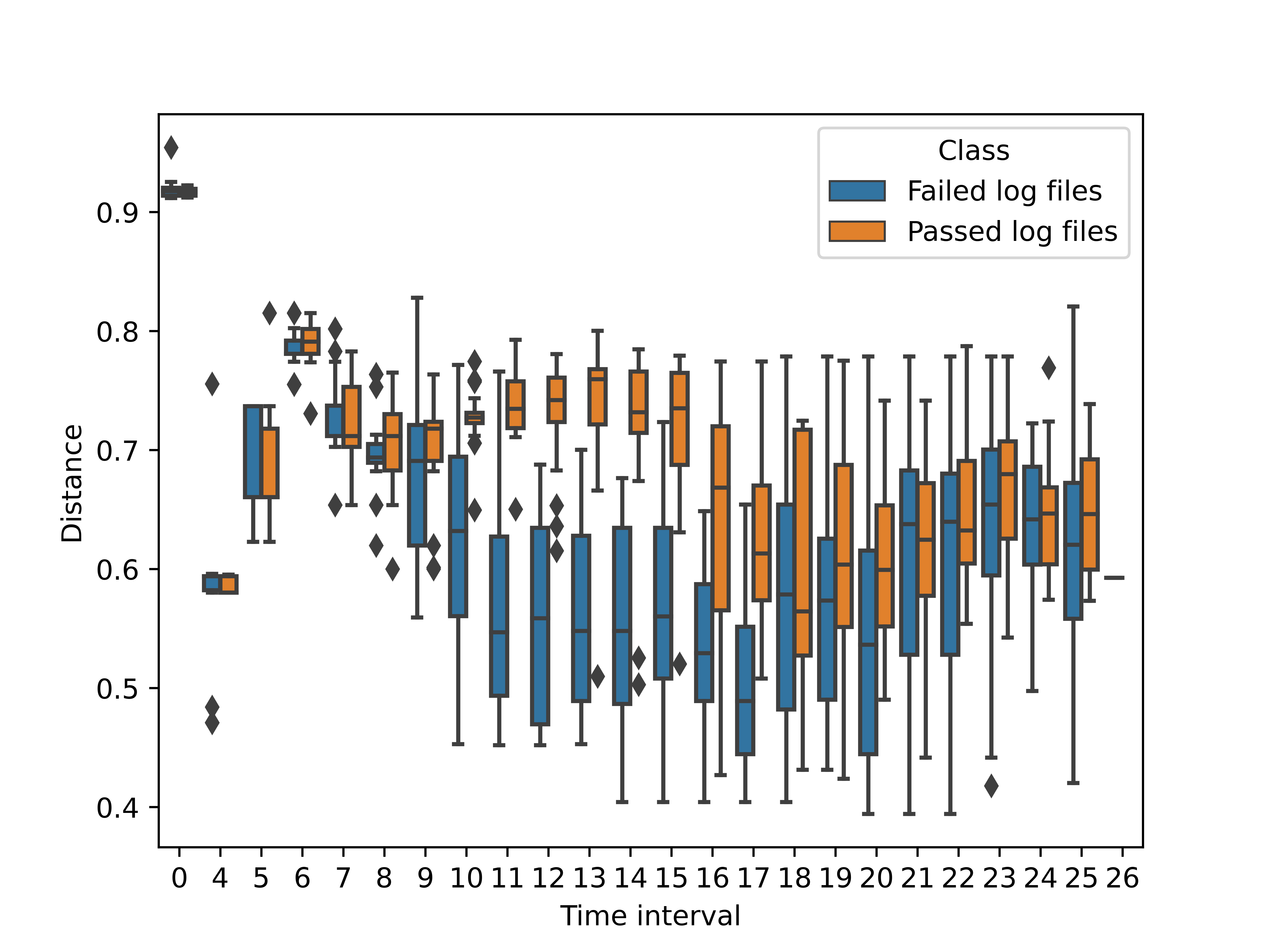}}
		\centerline{(a) Time interval from 0s to 25s}
	\end{minipage}
	\begin{minipage}{0.49\textwidth}
		\vspace{3pt}	\centerline{\includegraphics[width=\textwidth]{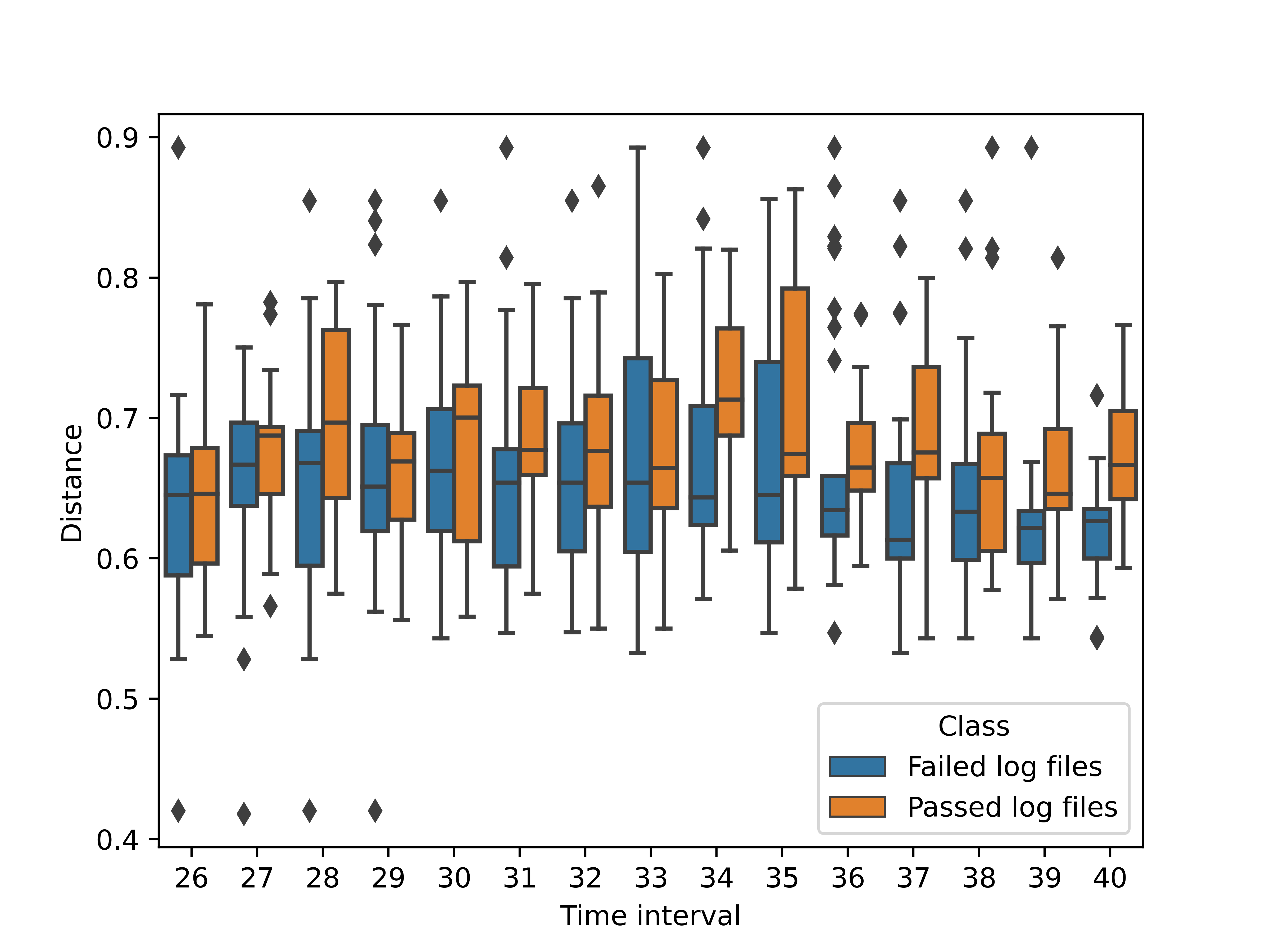}}
		\centerline{(b) Time interval from 26s to 40s}
	\end{minipage}
	\caption{The distribution of data features}
\end{figure}

\begin{figure}[htbp]
    \centering
    \centerline{\includegraphics[width=9cm]{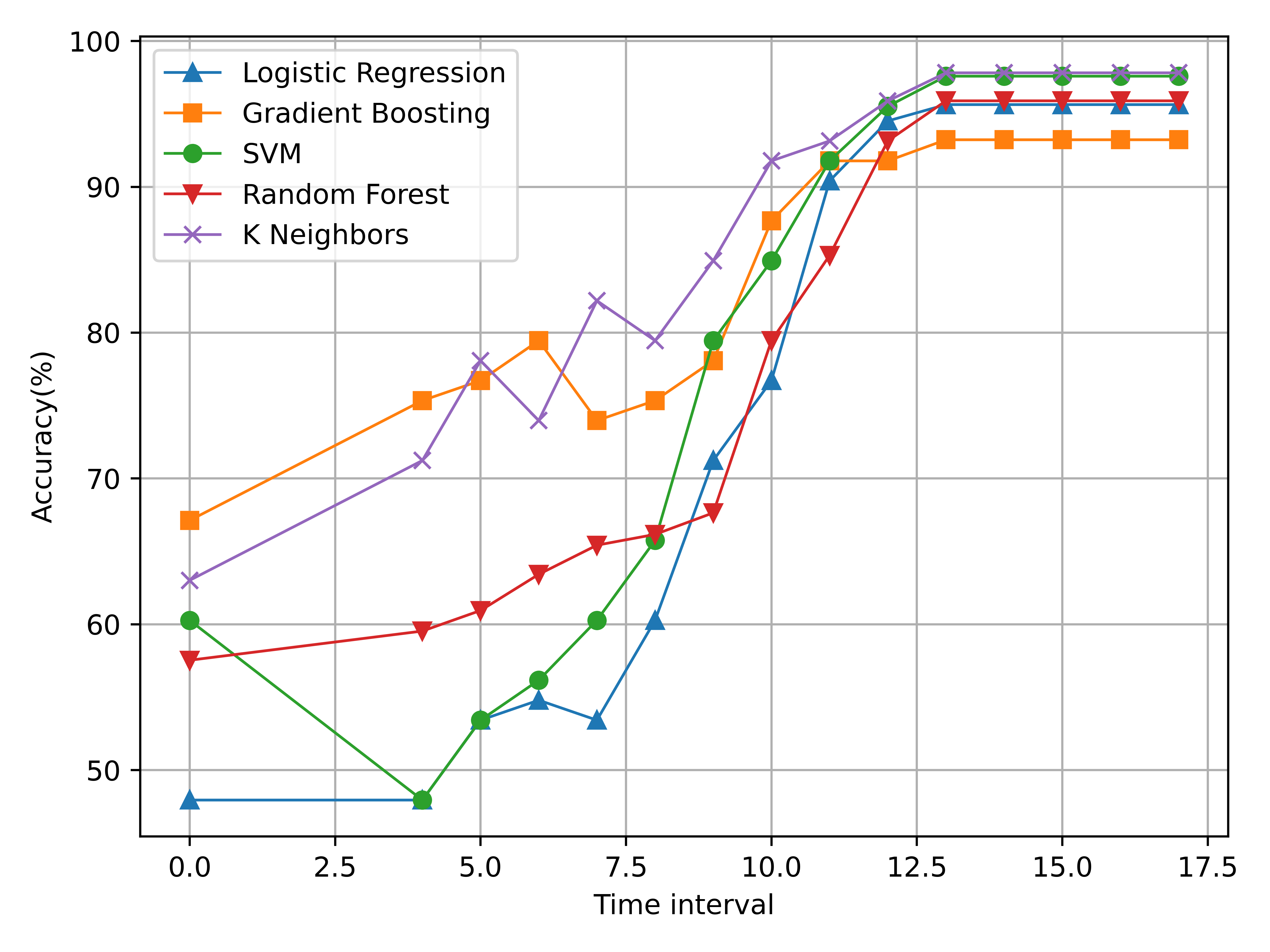}}
    \caption{Accuracy distribution changes at different time intervals}
\end{figure}

\begin{figure}[htbp]
    \centering
    \centerline{\includegraphics[width=9cm]{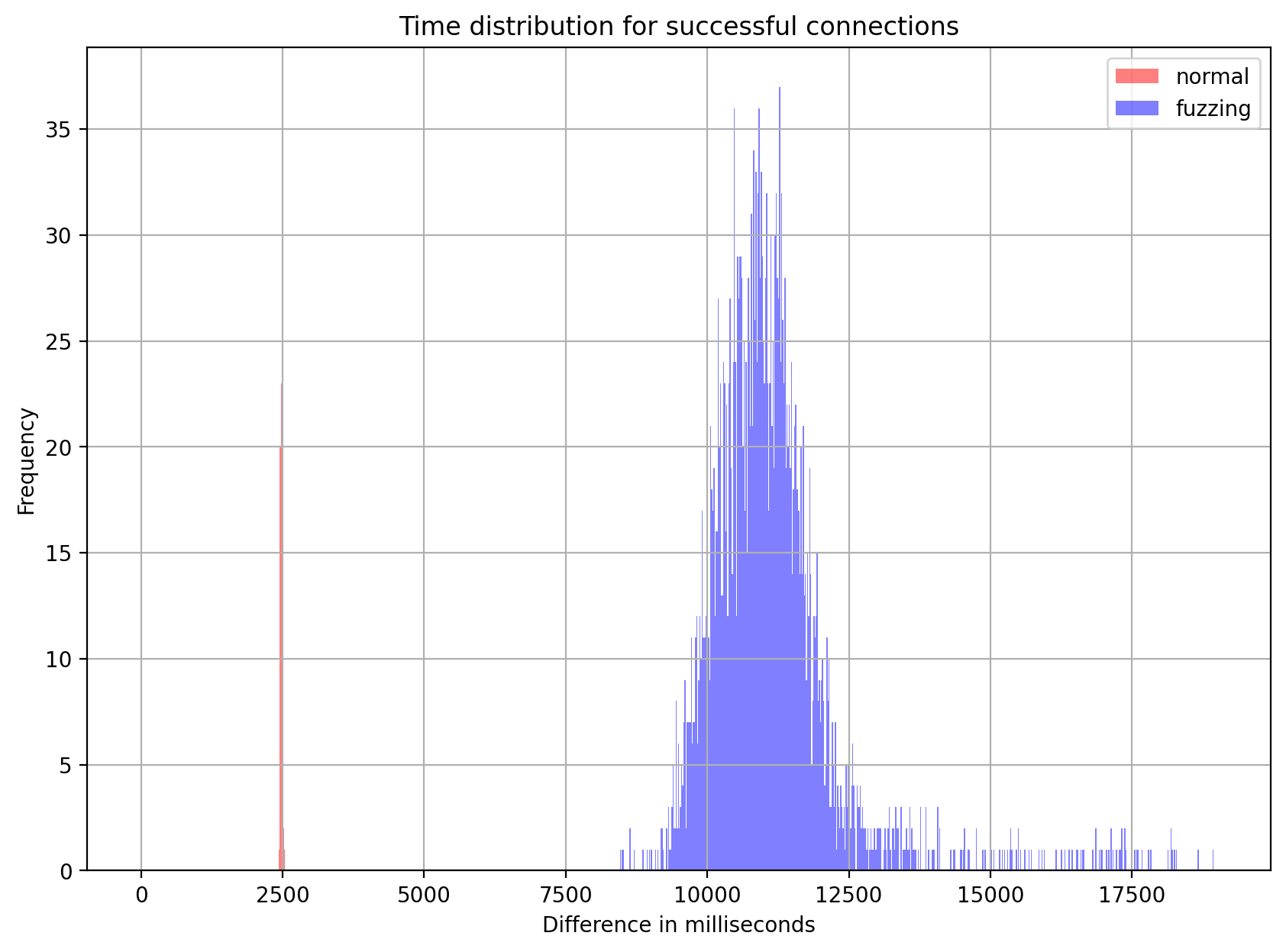}}
    \caption{Distribution of time to establish the connection between UE and ENB}
\end{figure}

\section{Conclusions and future work}
This paper studies the vulnerability detection method from an NLP perspective, aiming to provide a new solution idea. We have demonstrated the design and implementation of our method for real log files generated by fuzzing tests based on the 5G RRC protocol and achieved good practical performance. Detecting vulnerabilities automatically in 5G systems is difficult and complex work and our approach provides a new possible solution idea. 

As shown in Fig. 9, although these cases show the final result is connecting successfully, there are obvious and different delay situation that happens compared to the log files without any fuzzing injection. In real situations, some severe delays are equivalent to connection failures. And therefore, what we do next is to further analyze the relationship between fuzzing types and delay.

\bibliographystyle{ieeetr}
\bibliography{ref, references_ying} 

\begin{thebibliography}{10}

\bibitem{Hussain2019}
S.~R. Hussain, M.~Echeverria, I.~Karim, O.~Chowdhury, and E.~Bertino,
  ``5greasoner: A property-directed security and privacy analysis framework for
  5g cellular network protocol,'' pp.~669--684, Association for Computing
  Machinery, 11 2019.

\bibitem{9701880}
T.~Saha, N.~Aaraj, and N.~K. Jha, ``Machine learning assisted security analysis
  of 5g-network-connected systems,'' {\em IEEE Transactions on Emerging Topics
  in Computing}, vol.~10, no.~4, pp.~2006--2024, 2022.

\bibitem{9790271}
Y.~Cao, Y.~Chen, and W.~Zhou, ``Detection of pfcp protocol based on fuzz
  method,'' in {\em 2022 2nd International Conference on Computer, Control and
  Robotics (ICCCR)}, pp.~207--211, 2022.

\bibitem{9868872}
F.~He, W.~Yang, B.~Cui, and J.~Cui, ``Intelligent fuzzing algorithm for 5g nas
  protocol based on predefined rules,'' in {\em 2022 International Conference
  on Computer Communications and Networks (ICCCN)}, pp.~1--7, 2022.

\bibitem{9652163}
M.~Ajit, S.~Sankaran, and K.~Jain, ``Formal verification of 5g eap-aka
  protocol,'' in {\em 2021 31st International Telecommunication Networks and
  Applications Conference (ITNAC)}, pp.~140--146, 2021.

\bibitem{8923409}
J.~Zhang, Q.~Wang, L.~Yang, and T.~Feng, ``Formal verification of 5g-eap-tls
  authentication protocol,'' in {\em 2019 IEEE Fourth International Conference
  on Data Science in Cyberspace (DSC)}, pp.~503--509, 2019.

\bibitem{zhang2012research}
L.~Zhang, ``The research of log-based network monitoring system,'' in {\em
  Advances in intelligent systems}, pp.~315--320, Springer, 2012.

\bibitem{sipola2011anomaly}
T.~Sipola, A.~Juvonen, and J.~Lehtonen, ``Anomaly detection from network logs
  using diffusion maps,'' in {\em Engineering Applications of Neural Networks},
  pp.~172--181, Springer, 2011.

\bibitem{edris2020formal}
E.~K.~K. Edris, M.~Aiash, and J.~K.-K. Loo, ``Formal verification and analysis
  of primary authentication based on 5g-aka protocol,'' in {\em 2020 Seventh
  International Conference on Software Defined Systems (SDS)}, pp.~256--261,
  IEEE, 2020.

\bibitem{zhang2019formal}
J.~Zhang, Q.~Wang, L.~Yang, and T.~Feng, ``Formal verification of 5g-eap-tls
  authentication protocol,'' in {\em 2019 IEEE fourth international conference
  on data science in cyberspace (DSC)}, pp.~503--509, IEEE, 2019.

\bibitem{hu2021fuzzing}
Y.~Hu, W.~Yang, B.~Cui, X.~Zhou, Z.~Mao, and Y.~Wang, ``Fuzzing method based on
  selection mutation of partition weight table for 5g core network ngap
  protocol,'' in {\em International Conference on Innovative Mobile and
  Internet Services in Ubiquitous Computing}, pp.~144--155, Springer, 2021.

\bibitem{he2022intelligent}
F.~He, W.~Yang, B.~Cui, and J.~Cui, ``Intelligent fuzzing algorithm for 5g nas
  protocol based on predefined rules,'' in {\em 2022 International Conference
  on Computer Communications and Networks (ICCCN)}, pp.~1--7, IEEE, 2022.

\bibitem{9519388}
Y.~Chen, Y.~Yao, X.~Wang, D.~Xu, C.~Yue, X.~Liu, K.~Chen, H.~Tang, and B.~Liu,
  ``Bookworm game: Automatic discovery of lte vulnerabilities through
  documentation analysis,'' in {\em 2021 IEEE Symposium on Security and Privacy
  (SP)}, pp.~1197--1214, 2021.

\bibitem{10.1007/11736790_9}
I.~Dagan, O.~Glickman, and B.~Magnini, ``The pascal recognising textual
  entailment challenge,'' in {\em Machine Learning Challenges. Evaluating
  Predictive Uncertainty, Visual Object Classification, and Recognising Tectual
  Entailment} (J.~Qui{\~{n}}onero-Candela, I.~Dagan, B.~Magnini, and
  F.~d'Alch{\'e} Buc, eds.), (Berlin, Heidelberg), pp.~177--190, Springer
  Berlin Heidelberg, 2006.

\bibitem{10.1145/3372297.3423360}
T.~Lv, R.~Li, Y.~Yang, K.~Chen, X.~Liao, X.~Wang, P.~Hu, and L.~Xing, ``Rtfm!
  automatic assumption discovery and verification derivation from library
  document for api misuse detection,'' in {\em Proceedings of the 2020 ACM
  SIGSAC Conference on Computer and Communications Security}, CCS '20, (New
  York, NY, USA), p.~1837–1852, Association for Computing Machinery, 2020.

\bibitem{JingdaYang20235GListen-and-Learn}
{Jingda Yang}, {Ying Wang}, {Tuyen X. Tran}, and {Yanjun Pan}, ``{5G RRC
  Protocol and Stack Vulnerabilities Detection via Listen-and-Learn},'' in {\em
  IEEE Consumer Communications {\&} Networking Conference}, 2023.

\bibitem{SoftwareRadioSystems2021SrsRANSRS}
{Software Radio Systems}, ``{srsRAN is a 4G/5G software radio suite developed
  by SRS},'' 2021.

\bibitem{chen2021experience}
Z.~Chen, J.~Liu, W.~Gu, Y.~Su, and M.~R. Lyu, ``Experience report: deep
  learning-based system log analysis for anomaly detection,'' {\em arXiv
  preprint arXiv:2107.05908}, 2021.

\bibitem{devlin2018bert}
J.~Devlin, M.-W. Chang, K.~Lee, and K.~Toutanova, ``Bert: Pre-training of deep
  bidirectional transformers for language understanding,'' {\em arXiv preprint
  arXiv:1810.04805}, 2018.

\bibitem{reimers2019sentence}
N.~Reimers and I.~Gurevych, ``Sentence-bert: Sentence embeddings using siamese
  bert-networks,'' {\em arXiv preprint arXiv:1908.10084}, 2019.

\bibitem{reimers-2020-multilingual-sentence-bert}
N.~Reimers and I.~Gurevych, ``Making monolingual sentence embeddings
  multilingual using knowledge distillation,'' in {\em Proceedings of the 2020
  Conference on Empirical Methods in Natural Language Processing}, Association
  for Computational Linguistics, 11 2020.

\bibitem{van2008visualizing}
L.~Van~der Maaten and G.~Hinton, ``Visualizing data using t-sne.,'' {\em
  Journal of machine learning research}, vol.~9, no.~11, 2008.

\end{thebibliography}
\end{document}